\documentclass[twocolumn,english,aps,manuscript,onecolumn,superscriptaddress]{revtex4}
\usepackage[LGR,LGR,T1]{fontenc}
\usepackage[latin9]{inputenc}
\usepackage{units}
\usepackage{textcomp}
\usepackage{amsmath}
\usepackage{amssymb}
\usepackage{graphicx}
\usepackage{esint}

\makeatletter

\newcommand*\LyXThinSpace{\,\hspace{0pt}}
\DeclareRobustCommand{\greektext}{%
  \fontencoding{LGR}\selectfont\def\encodingdefault{LGR}}
\DeclareRobustCommand{\textgreek}[1]{\leavevmode{\greektext #1}}
\ProvideTextCommand{\~}{LGR}[1]{\char126#1}

\@ifundefined{textcolor}{}
{%
 \definecolor{BLACK}{gray}{0}
 \definecolor{WHITE}{gray}{1}
 \definecolor{RED}{rgb}{1,0,0}
 \definecolor{GREEN}{rgb}{0,1,0}
 \definecolor{BLUE}{rgb}{0,0,1}
 \definecolor{CYAN}{cmyk}{1,0,0,0}
 \definecolor{MAGENTA}{cmyk}{0,1,0,0}
 \definecolor{YELLOW}{cmyk}{0,0,1,0}
}

\usepackage[english]{babel}
\def\urlprefix{}
\def\url#1{}
\addto\captionsenglish{%
}

\usepackage[none]{hyphenat}

\usepackage{babel}

\usepackage{babel}

\makeatother

\usepackage{babel}
\begin{document}

\title{Topological transitions induced by antiferromagnetism in a thin-film
topological insulator}

\author{Qing Lin He\footnotemark[1]\footnotemark[2]}

\affiliation{Department of Electrical Engineering, University of California, Los
Angeles, California 90095, USA.}

\author{Gen Yin}

\thanks{These authors contributed to this work equally.}

\affiliation{Department of Electrical Engineering, University of California, Los
Angeles, California 90095, USA.}

\author{Luyan Yu}

\thanks{These authors contributed to this work equally.}

\affiliation{Department of Electrical Engineering, University of California, Los
Angeles, California 90095, USA.}

\author{Alexander J. Grutter}

\affiliation{NIST Center for Neutron Research, National Institute of Standards
and Technology, Gaithersburg, MD 20899-6102, USA.}

\author{Lei Pan}

\affiliation{Department of Electrical Engineering, University of California, Los
Angeles, California 90095, USA.}

\author{Xufeng Kou}

\thanks{Correspondence to: qlhe@ucla.edu; kouxf@shanghaitech.edu.cn; wang@ee.ucla.edu.}

\affiliation{School of Information Science and Technology, ShanghaiTech University,
Shanghai 200031, China.}

\author{Xiaoyu Che}

\affiliation{Department of Electrical Engineering, University of California, Los
Angeles, California 90095, USA.}

\author{Guoqiang Yu}

\affiliation{Department of Electrical Engineering, University of California, Los
Angeles, California 90095, USA.}

\author{Tianxiao Nie}

\affiliation{Department of Electrical Engineering, University of California, Los
Angeles, California 90095, USA.}

\author{Bin Zhang}

\affiliation{Beijing Key Lab of Microstructure and Property of Advanced Materials,
Beijing University of Technology, 100124, Beijing, China.}

\author{Qiming Shao}

\affiliation{Department of Electrical Engineering, University of California, Los
Angeles, California 90095, USA.}

\author{Koichi Murata}

\affiliation{Department of Electrical Engineering, University of California, Los
Angeles, California 90095, USA.}

\author{Xiaodan Zhu}

\affiliation{Department of Electrical Engineering, University of California, Los
Angeles, California 90095, USA.}

\author{Yabin Fan}

\affiliation{Department of Electrical Engineering, University of California, Los
Angeles, California 90095, USA.}

\author{Xiaodong Han}

\affiliation{Beijing Key Lab of Microstructure and Property of Advanced Materials,
Beijing University of Technology, 100124, Beijing, China.}

\author{Brian J. Kirby}

\affiliation{NIST Center for Neutron Research, National Institute of Standards
and Technology, Gaithersburg, MD 20899-6102, USA.}

\author{Kang L. Wang}

\thanks{Correspondence to: qlhe@ucla.edu; kouxf@shanghaitech.edu.cn; wang@ee.ucla.edu.}

\affiliation{Department of Electrical Engineering, University of California, Los
Angeles, California 90095, USA.}
\begin{abstract}
Ferromagnetism in topological insulators (TIs) opens a topologically
non-trivial exchange band gap, providing an exciting platform to manipulate
the topological order through an external magnetic field. Here, we
experimentally show that the surface of an antiferromagnetic thin
film can independently control the topological order of the top and
the bottom surface states of a TI thin film through proximity couplings.
During the magnetization reversal in a field scan, two intermediate
spin configurations stem from unsynchronized magnetic switchings of
the top and the bottom AFM/TI interfaces. These magnetic configurations
are shown to result in new topological phases with non-zero Chern
numbers for each surface, introducing two counter-propagating chiral
edge modes inside the exchange gap. This change in the number of transport
channels, as the result of the topological transitions, induces antisymmetric
magneto-resistance spikes during the magnetization reversal. With
the high N�el ordering temperature provided by the antiferromagnetic
layers, the signature of the induced topological transition persists
in transport measurements up to a temperature of around $90\thinspace\textrm{K}$,
a factor of three over the Curie temperature in a typical magnetically
doped TI thin film. 
\end{abstract}
\maketitle
Currently there is immense interest in the manipulation of ferromagnetic
phases in topological insulators (TIs) through either doping with
magnetic species or through proximity coupling to a strong ferromagnetic
system. This interest is driven by the novel physics which is predicted
to emerge as a consequence of the TIs' non-trivial topology in $k$
space \cite{chang_experimental_2013,kou_scale-invariant_2014,yu_quantized_2010}.
The Dirac fermion surface states of three-dimensional TIs are robust
against lattice defect perturbations, non-magnetic imperfections,
and surface reconstructions \cite{qi_topological_2011,hasan_textitcolloquium_2010,kane_$z_2$_2005,zhang_topological_2009,bernevig_quantum_2006}.
Breaking time-reversal symmetry in these systems with magnetic dopants
such as Cr or V opens an exchange band gap, inducing a finite Berry
curvature and leading to an intrinsic anomalous Hall effect (AHE)
\cite{nagaosa_anomalous_2010,xiao_berry_2010}. Inside this exchange
gap, non-zero Chern numbers of $\pm1$ arise, protecting a chiral
AHE edge mode which encircles the boundary of the TI thin film \cite{chang_experimental_2013,chang_high-precision_2015,kou_scale-invariant_2014,yu_quantized_2010}.
Similar to the edge mode in conventional quantum Hall effects, back
scattering is forbidden in the chiral mode, enabling dissipationless
charge transport. Furthermore, the direction of the momentum and spin
of the surface Dirac fermions are locked to be perpendicular, leading
to many potential spintronic applications \cite{fan_magnetization_2014,mellnik_spin-transfer_2014}.
These unique features make TIs a distinctive and exciting platform
for investigating the interplay between charge and spin in topologically
non-trivial matters.

Despite their exotic magneto-transport properties, the implementation
of novel spintronic devices based on TI systems remain conceptual,
largely due to the low magnetic ordering temperatures, small exchange
gaps, and the potential fluctuations induced by the randomly distributed
dopants. In search for high-temperature TI spintronics, conventional
mechanisms to introduce ferromagnetic order suffer from significant
limitations. For example, although higher Cr and V-doping levels increase
the magnitude of the exchange gap, they also modify the band order
such that the topologically protected surface states may be eliminated
in heavily doped magnetic TIs \cite{kou_magnetically_2012,zhang_topology-driven_2013,jiang_mass_2015}.
This confinement to low-level doping restricts accessible Curie temperatures
($T_{\textrm{C}}$) to cryogenic temperatures, as the large distance
between magnetic dopants weakens the Heisenberg exchange coupling
between neighbouring spins. The other common method for achieving
ferromagnetic TIs is interfacial proximity coupling to adjacent ferromagnets.
Since the magnetic order in this case is introduced extrinsically
by an adjacent ferromagnetic layer with a high $T_{\textrm{C}}$,
it may persist at much higher temperatures than that induced by magnetic
doping \cite{lang_proximity_2014,jiang_independent_2015,katmis_high-temperature_2016}.
Proximity induced magnetization of the TI in these systems is both
hard to achieve and extremely difficult to characterize, with the
signal from the ferromagnet swamping out the contribution from the
TI layers. Furthermore, the ferromagnet produces significant stray
fields and is easily affected by perturbations in external magnetic
fields or currents, making crosstalk a problem, adversely affecting
the device scalability \cite{jungwirth_antiferromagnetic_2016}.

Here, we propose the use of an antiferromagnet (AFM) interfacing to
a TI thin film as an alternative method to introduce a higher-$T_{\textrm{C}}$
magnetic order. Similar to the ferromagnet/TI system, the proximity-induced
magnetization in the AFM/TI structure originates in the localized
electronic wave-function overlap at the interface owing to the magnetically
uncompensated surface of the AFM \cite{luo_massive_2013}. More importantly,
proximity to AFMs has been experimentally shown to enhance the surface
magnetization of a TI thin film induced by the interfacial exchange
coupling \cite{he_tailoring_2016-1}. AFMs have vanishingly small
net magnetization and consequently neither produce stray fields nor
affect the characterization of the TI layer. Therefore the magnetic
order is robust against the external magnetic field or moderate current
perturbations, minimizing the crosstalk between devices and improving
scalability. Further, the precession of one spin sub-lattice in an
AFM experiences a comparatively large effective field via the Heisenberg
exchange coupling to the other spin sub-lattice \cite{keffer_theory_1952}.
Consequently, AFM spintronic devices can reach operating frequencies
in the THz range, much higher than the GHz range accessible in ferromagnet-based
devices \cite{cheng_terahertz_2016,jungwirth_antiferromagnetic_2016}.
In this letter, we experimentally demonstrate the proximity-induced
ferromagnetism at the AFM/TI {[}$\textrm{CrSb/(Bi,Sb)}_{2}\textrm{Te}_{3}${]}
interface in an AFM/TI bilayer, an AFM/TI/AFM trilayer, and an $\textrm{[AFM/TI]}_{\textrm{n}}$
superlattice system. The uncompensated local spins at the interface
magnetize the surface Dirac fermions in the TI robustly, giving rise
to an observed AHE that survives up to 90 K, beyond the $T_{\textrm{C}}$
of a typical magnetically doped TI of \textasciitilde{} 30 K. Equally
important, we observed that the magnetizations of the top and the
bottom TI surfaces can switch in a step-by-step manner. As a result,
two antisymmetric magnetoresistance (MR) peaks show up during the
sweep of the magnetic field. These two peaks correspond to different
configurations in the spin texture, representing two distinct topological
orders. 
\begin{figure}
\begin{centering}
\includegraphics[width=0.9\columnwidth]{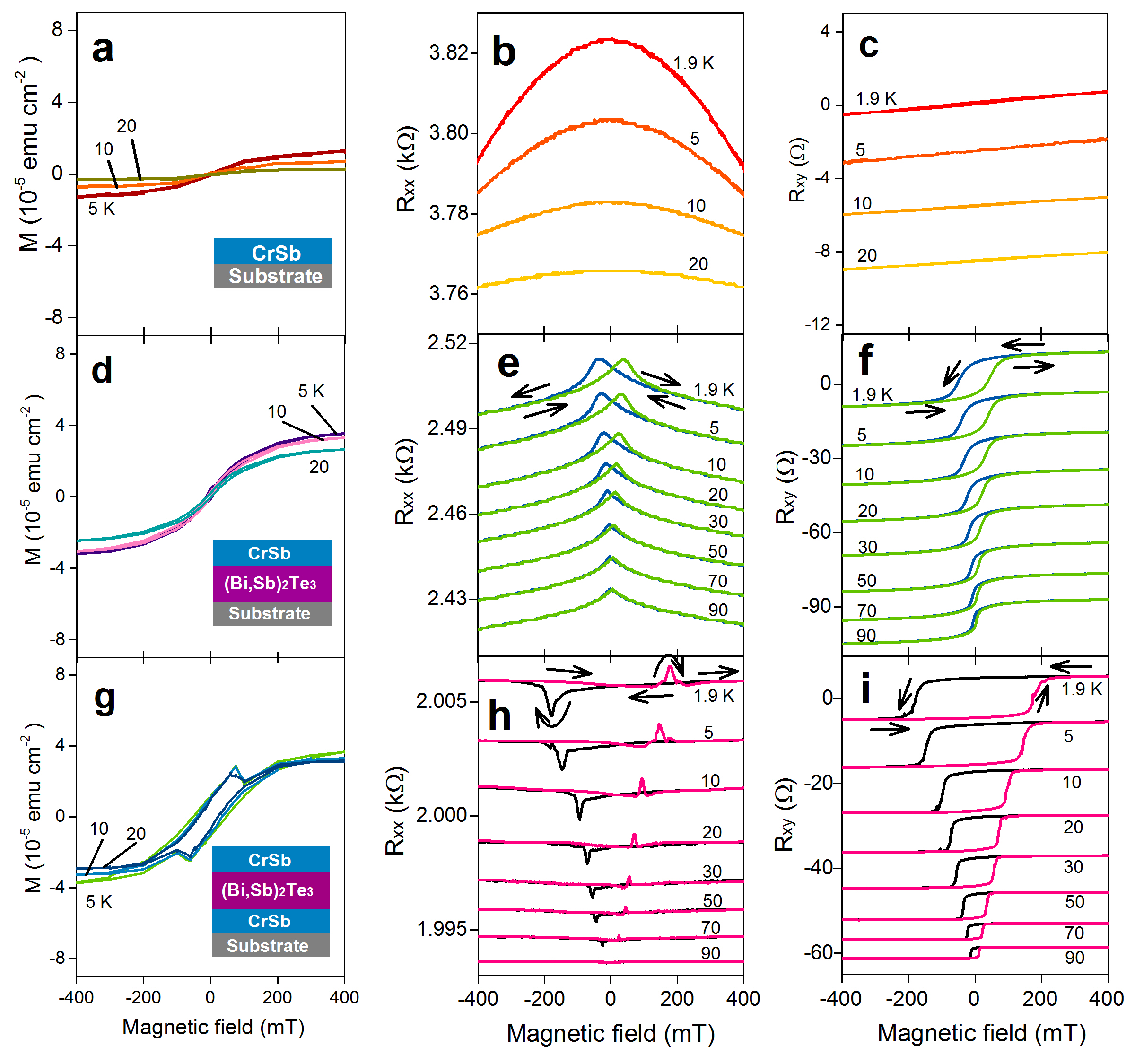} 
\par\end{centering}
\caption{\textbf{Emerging surface magnetizations in antiferromagnet/topological
insulator (AFM/TI) structures.} \textbf{a}, temperature-dependent
M-H loops of a 12-nm AFM thin film, CrSb, grown on a GaAs substrate,
which exhibits negligible magnetization. \textbf{b} and \textbf{c}
show the results of the longitudinal ($R_{xx}$) and the Hall ($R_{xy}$)
resistance, respectively. The absence of hysteresis in both measurement
results, i.e. the small parabolic curvature in $R_{xx}$ and the linear
response in $R_{xy}$, suggest that the CrSb layer does not generate
an AHE intrinsically. \textbf{d}-\textbf{f} show the corresponding
results of an AFM/TI bilayer, demonstrating the transport signature
of the top-surface magnetization of the TI layer induced by proximity.
\textbf{g}-\textbf{i} are from an AFM/TI/AFM trilayer, in which an
antisymmetric MR behavior and an AHE are observed. Compared with the
bilayer case, the trilayer contains two individually magnetized TI
surfaces due to the proximity to two CrSb layers, respectively. This
allows an intermediate topological order of the TI thin film when
the top and bottom TI surfaces switch in a step-by-step manner. ($1\thinspace\textrm{\ensuremath{\mu}emu}=1\thinspace\textrm{nA}\cdot\textrm{m}^{2}$)
\label{fig:SQUID-and-transport}}
\end{figure}

Firstly, in order to ascertain the magnetic order in CrSb thin films,
superconducting quantum interference device magnetometry (SQUID) and
magneto-transport measurements were carried out on a $12\thinspace\textrm{nm}$
CrSb layer grown on a GaAs substrate. An external magnetic field was
applied perpendicularly to the film plane. As shown in Fig. \ref{fig:SQUID-and-transport}a,
a vanishingly weak magnetization is captured in the M-H measurement,
diminishing quickly as the temperature increases to $\sim20\,\textrm{K}$.
Such a small magnetization is likely to originate from the uncompensated
spins in the AFM layer at the interface \cite{luo_massive_2013,he_tailoring_2016-1}.
Consistently, magneto-transport measurements also capture a small
negative MR which also vanishes at around $20\thinspace\textrm{K}$
(Fig. \ref{fig:SQUID-and-transport}b). The corresponding Hall resistance
obtained (Fig. \ref{fig:SQUID-and-transport}c) in a $2\thinspace\textrm{mm}\times1\thinspace\textrm{mm}$
Hall bar varies linearly with the external field, indicating that
the Hall signal is mainly associated with the ordinary Hall effect.
These results suggest that the weak magnetization in the CrSb thin
film results in neither magnetic hysteresis nor intrinsic AHE.

Strikingly, when the AFM layer was epitaxially grown on a $8\textrm{-nm}$
TI thin film, a proximity-induced AHE was observed. As shown in Fig.
\ref{fig:SQUID-and-transport}d, the saturation magnetization is roughly
$4\times10^{-5}\,\textrm{emu}\cdot\textrm{cm}^{-2}$, which is higher
than that of the single AFM layer grown on the substrate (Fig. 1a).
This indicates that the polarized TI surface spins contribute extra
magnetic moments to the saturated magnetization. Furthermore, the
MR measurement exhibits a hysteretic butterfly shape which gradually
vanishes as the temperature increases to 90 K, as shown in Fig. \ref{fig:SQUID-and-transport}e.
The corresponding Hall resistance results are shown in Fig. \ref{fig:SQUID-and-transport}f.
In this figure, a significant AHE signal is evidenced by the large
hysteresis even without subtracting the linear ordinary Hall resistance,
indicating a strong magnetic proximity effect in the AFM/TI bilayer.
Moreover, the observed AHE persists at 90 K, which is roughly a factor
of three higher than the $T_{\textrm{C}}$ of a typical Cr-doped TI.
This enhancement in $T_{\textrm{C}}$ implies that the AHE does not
originate from Cr diffusion into the TI layer, but is rather dominated
by a proximity-induced AHE, since an extremely heavy Cr doping concentration
would be required to achieve such a high $T_{\textrm{C}}$. In such
a heavily-doped case, the TI film would exhibit a strong perpendicular
magnetization, sharply contradicted by the weak magnetic moment observed
in the AFM/TI bilayer with SQUID, as shown in Fig. \ref{fig:SQUID-and-transport}d. 

Surprisingly, when introducing one additional AFM layer to produce
a AFM/TI/AFM sandwich structure, another distinct magnetic feature
is revealed. In contrast to Figs. 1a and d, the M-H results show clear
hysteresis loops from 5 K to 20 K (Fig. \ref{fig:SQUID-and-transport}g),
which implies the presence of unsynchronized switching of the top
and the bottom AFM layers, given that the interfacial exchange coupling
experienced by the uncompensated spins is strongly determined by the
details (e.g. roughness, defect density, etc.) at the interfaces.
Corresponding to this double-switching, the temperature-dependent
MR curves show antisymmetric spikes during the magnetization reversal
process, as shown in Fig. \ref{fig:SQUID-and-transport}h, and these
spikes occur when the corresponding AHE resistance changes sign (Fig.
\ref{fig:SQUID-and-transport}i). Meanwhile, both the spike-to-spike
distance of the MR curve and the coercivity of the AHE loop have a
temperature dependence which closely matches that of the AFM/TI case
with a phase change temperature of around 90 K, underscoring the importance
of the interfacial magnetic proximity effect induced by the two AFM
layers. This antisymmetric MR contrasts sharply with the symmetric
MR observed in most ferromagnetic systems \cite{lang_proximity_2014},
as well as the AFM/TI bilayer in this work. We note that all the M-H,
MR, and AHE results presented in Fig. \ref{fig:SQUID-and-transport}
are reproducible across different batches of samples/devices, enabling
us to exclude experimental artifacts related to the device architectures
(e.g. misalignment of the Hall voltage leads, device sizing effects,
etc.). Consequently, we propose that the observed antisymmetric MR
of the AFM/TI/AFM trilayer samples has a surprising origin - topological
transitions induced by the double-switching mechanism, as discussed
below.

\begin{figure*}
\begin{centering}
\includegraphics[width=0.9\textwidth]{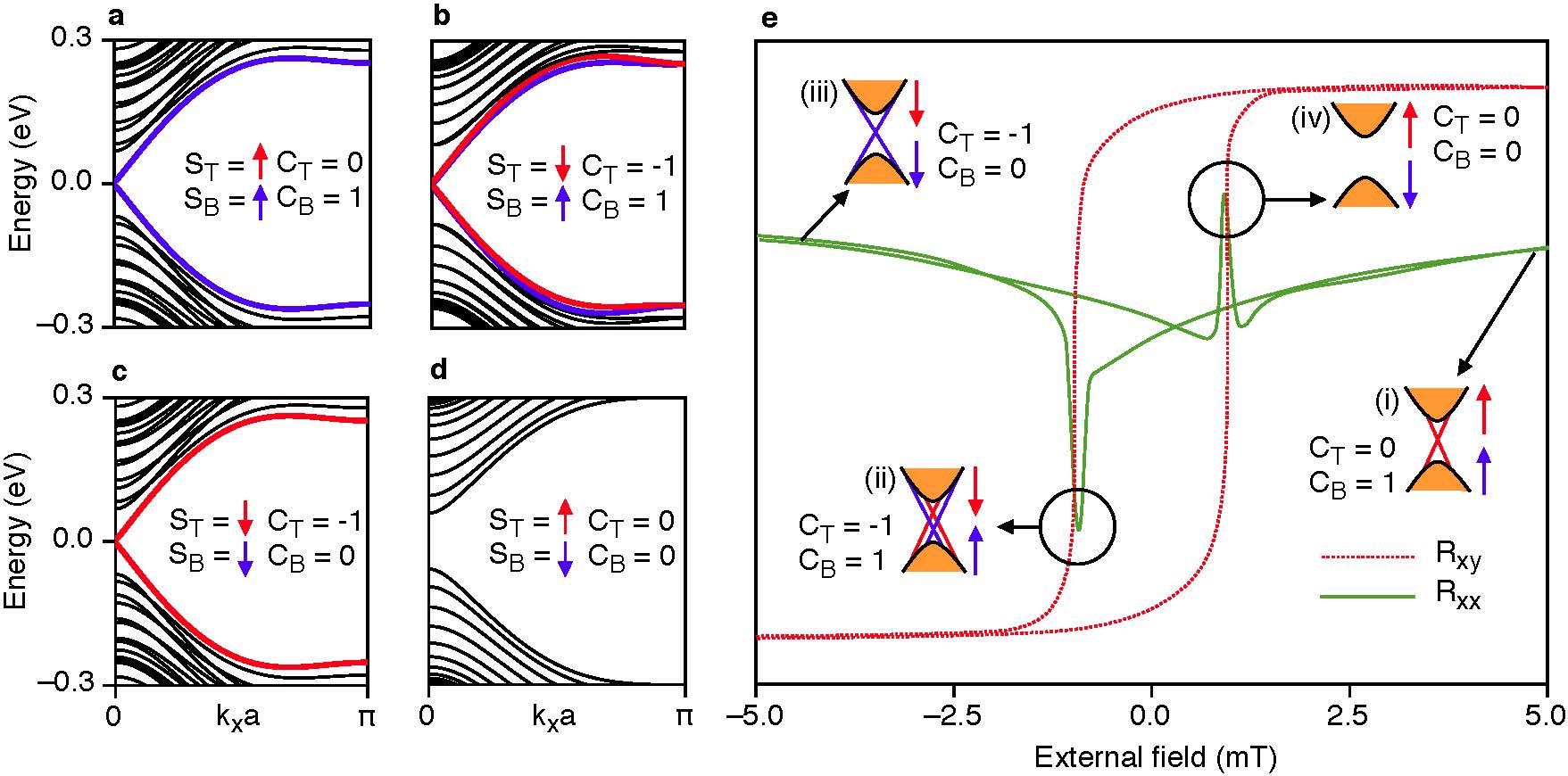} 
\par\end{centering}
\caption{\textbf{The topological transition during the double-switching magnetization
reversal.} \textbf{a}-\textbf{d}, The band structure of a TI thin
film with different top-bottom (red-blue) spin configurations ($B=16.45\thinspace\textrm{eV�}^{-2}$,
$\gamma=3.29\thinspace\textrm{eV�}^{-1}$, $J_{\textrm{H}}=0.05\thinspace\textrm{eV}$).
A negligibly small top-bottom potential drop is numerically added
to \textbf{b} for clarity. \textbf{e}, The topological transition
during the reversal of the total magnetization as remarked near the
AHE hysteresis and the corresponding MR curves. The four different
topological orders (i-iv) and their band structures are also schematically
shown at different transition points during the field scan. These
four different cases (i-iv) correspond to the band structures as shown
in \textbf{a}-\textbf{d}. \label{fig:The-theoretical-figure2}}
\end{figure*}

In the AFM/TI/AFM trilayer, the TI layer can be described by the effective
four-band Hamiltonian \cite{shan_effective_2010}:

\begin{equation}
H_{0}=\left(\begin{array}{cccc}
-Bk^{2} & i\gamma k_{-} & V & 0\\
-i\gamma k_{+} & Bk^{2} & 0 & V\\
V & 0 & Bk^{2} & i\gamma k_{-}\\
0 & V & -i\gamma k_{+} & -Bk^{2}
\end{array}\right)+H_{\textrm{H}}\label{eq:K_dot_P_Hamiltonian}
\end{equation}
where $B>0$ is the parabolic massive component induced by the surface-to-surface
coupling, $V$ denotes the structural inversion asymmetry due to the
top-to-bottom electrostatic potential drop, and $\gamma=\hbar v_{F}$,
where $v_{F}$ is the constant group velocity of the surface Dirac
fermions \cite{lu_quantum_2013,lu_massive_2010}. Due to the surface
magnetization, a Hund's-rule coupling term, $H_{\textrm{H}}$, is
added to break the time-reversal symmetry: 
\begin{equation}
H_{\textrm{H}}=-J_{\textrm{H}}\left(\begin{array}{cc}
\boldsymbol{S}_{\textrm{T}}\cdot\boldsymbol{\sigma} & 0\\
0 & \boldsymbol{S}_{\textrm{B}}\cdot\boldsymbol{\sigma}
\end{array}\right)\label{eq:HundsRuleCoupling}
\end{equation}
Here, $\boldsymbol{S}_{\textrm{T}}$ and $\boldsymbol{S}_{\textrm{B}}$
represent the spin configurations of the top and the bottom surface
Dirac fermions under the influence of the AFM-induced proximity effect.
$J_{\textrm{H}}$ denotes the interfacial Hund's-rule coupling and
in this case $J_{\textrm{H}}>0$. In order to model the magneto-electric
transport behavior of the trilayer structure, we establish a tight-binding
model by discretizing $k_{\alpha}\rightarrow-i\partial_{\alpha}$.
Usually applying a Dirac fermion Hamiltonian in a lattice structure
introduces a Fermi doubling problem, such that the dispersion unphysically
closes the gap at the boundary of the Brillouin zone \cite{stacey_eliminating_1982,susskind_lattice_1977}.
However, in the thin-film case, the non-zero surface-to-surface hybridization-induced
parabolic term resolves this problem \cite{habib_modified_2016}.
The band structure of a thin film with confined boundaries can be
obtained using this tight binding model with different spin configurations,
as shown in Figs. \ref{fig:The-theoretical-figure2}a-d. Note that
$H_{\textrm{H}}$ adds massive terms to the total Hamiltonian, such
that the reversals of $\boldsymbol{S}_{\textrm{T}}$ and $\boldsymbol{S}_{\textrm{B}}$
change the band order, inducing topological transitions \cite{qi_chiral_2010}.
Assuming a negligible structural inversion asymmetry term, $V=0$.
Thus, the two surfaces are block-diagonalized such that the total
Chern number is the sum of the Chern numbers given by the two surfaces:
$C=C_{\textrm{T}}+C_{\textrm{B}}$. Unlike the case of a pure Dirac
fermion, due to the surface-to-surface hybridization, a finite $B>0$
in Eq. \ref{eq:K_dot_P_Hamiltonian} makes the upper-left block a
typical $2\times2$ Hamiltonian of the quantum AHE, where integer
Chern numbers of $\pm1$ are introduced inside the exchange gap, rather
than $\pm\nicefrac{1}{2}$ \cite{qi_chiral_2010,shan_effective_2010}.
As shown in Figs. \ref{fig:The-theoretical-figure2}a-d, the four
different spin configurations correspond to different combinations
of $C_{\textrm{T}}$ and $C_{\textrm{B}}$. When $\boldsymbol{S}_{\textrm{T}}$
and $\boldsymbol{S}_{\textrm{B}}$ are parallel, integer total Chern
numbers are introduced: $C=\pm1$ (Figs. \ref{fig:The-theoretical-figure2}a
and c). In an ideal case where only the TI surface states contribute
to the conduction, these two non-zero Chern number configurations
give rise to the quantum AHE, where boundary modes of opposite chiralities
are established with different saturation directions of the total
magnetization. The dispersions of these modes are shown as the coloured
linear bands in Figs. \ref{fig:The-theoretical-figure2}a and c. 

During the reversal of the magnetization, unsynchronized switching
occurs, inducing two intermediate spin configurations with different
topological orders. On one hand, when $\boldsymbol{S}_{\textrm{T}}=\thinspace\uparrow$
and $\boldsymbol{S}_{\textrm{B}}=\thinspace\downarrow$, $C_{\textrm{T}}=C_{\textrm{B}}=0$,
and the magnetic exchange gap in the TI becomes trivial (Fig. \ref{fig:The-theoretical-figure2}d).
On the other hand, in the case when $\boldsymbol{S}_{\textrm{T}}=\thinspace\downarrow$
and $\boldsymbol{S}_{\textrm{B}}=\thinspace\uparrow$, the non-trivial
edge modes with two-fold degeneracy appear (i.e., $C_{\textrm{T}}=-1$
and $C_{\textrm{B}}=1$, see Fig. \ref{fig:The-theoretical-figure2}b),
yet the total Chern number is still $C=C_{\textrm{T}}+C_{\textrm{B}}=0$.
Accordingly, in both of the two configurations (Figs. \ref{fig:The-theoretical-figure2}b
and d), the Hamiltonian in Eq. \ref{eq:K_dot_P_Hamiltonian} restores
its time-reversal symmetry, such that the total Berry curvature is
zero and the intrinsic Hall signal is suppressed. However, since the
number of modes inside the exchange gap changes by two, the MR changes
significantly when the Fermi level ($\epsilon_{F}$) is close to the
Dirac point, leading to the antisymmetric $R_{xx}$ behavior. In Fig.
\ref{fig:The-theoretical-figure2}e, typical $R_{xx}$ and $R_{xy}$
of an AFM/TI/AFM trilayer are plotted together, where band structures
with different topological orders are also illustrated at different
stages of the magnetic field scan.

\begin{figure}
\begin{centering}
\includegraphics[width=0.8\columnwidth]{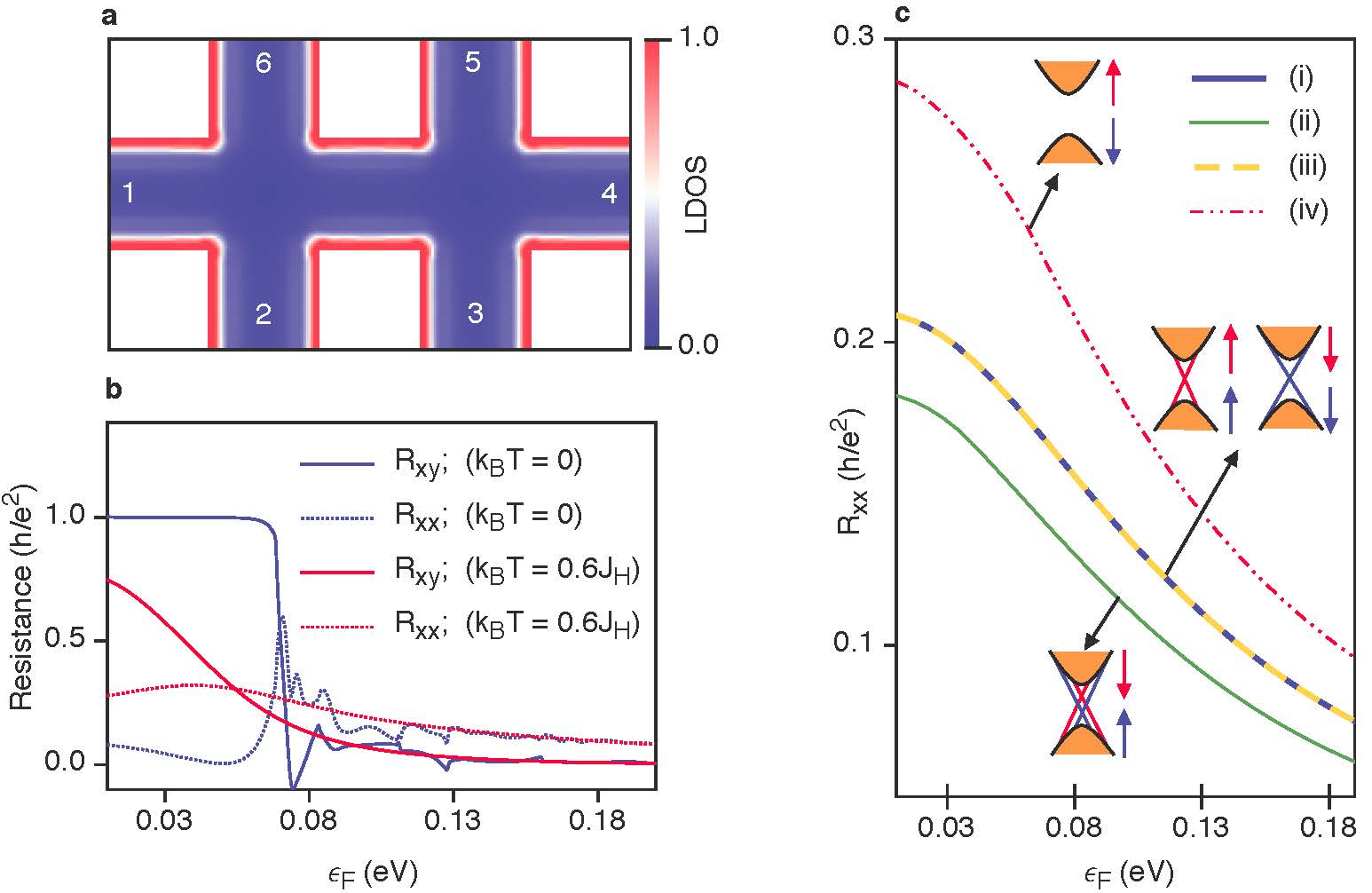}
\par\end{centering}
\caption{\textbf{Numerical simulation results of the charge transport in the
Hall bar.} \textbf{a}, The device structure constructed by the tight-binding
model corresponding to Eq. \ref{eq:K_dot_P_Hamiltonian}.\textbf{
}Local density of states calculated at $\epsilon_{\textrm{F}}=0.03\thinspace\textrm{eV}$
is illustrated by the colour plot. \textbf{b}, $R_{xy}$ and $R_{xx}$
obtained from a nonequilibrium Green's function calculation at different
temperatures. Longitudinal current is applied between terminal 1 and
4. $R_{xy}$ is obtained from terminal $2$ and $6$, while $R_{xx}$
is calculated between terminal $2$ and $3$. \textbf{c}, $R_{xx}$
as a function of $\epsilon_{F}$ for the four different spin configurations
($k_{\textrm{B}}T=1.4J_{\textrm{H}}$). Details of the band structures
corresponding to different topological orders (i-iv) are schematically
shown for each case. Note that the yellow and blue lines coincide
together. \label{fig:Numerical-simulation-results}}
\end{figure}

In order to numerically evaluate the longitudinal resistance, the
transport behavior of a six-terminal Hall bar (as shown in Fig. \ref{fig:Numerical-simulation-results}a)
is numerically evaluated using nonequilibrium Green's functions, based
on the tight-binding Hamiltonian established from Eq. \ref{eq:K_dot_P_Hamiltonian}
(see Methods) \cite{datta_quantum_2005,lake_single_1997}. At zero
temperature, the Hall resistance, $R_{xy}$, and the longitudinal
resistance, $R_{xx}$ as functions of $\epsilon_{F}$ are shown in
Fig. \ref{fig:Numerical-simulation-results}b, corresponding to the
case of $\boldsymbol{S}_{\textrm{T}}=\mathbf{S}_{\textrm{B}}=\thinspace\uparrow$.
A quantized plateau of $R_{xy}$ and a zero $R_{xx}$ are demonstrated
when $\epsilon_{F}$ is within the exchange gap. The local density
of states near the Dirac point of the Hall bar is shown as a colour
contrast in Fig. \ref{fig:Numerical-simulation-results}a, where a
single state of the chiral mode is localized at the boundaries. When
$k_{B}T$ increases to $0.6J_{\textrm{H}}$, the thermal broadening
begins to include sub-bands, such that the quantization of the AHE
begins to vanish, namely, $R_{xy}$ decreases while $R_{xx}$ increases
inside the exchange gap. In the experiment, the AHE does not reach
the quantized level. This indicates that states outside the magnetic
exchange band gap are involved in the transport due to either the
thermal broadening or the band-edge fluctuations. In this calculation,
$k_{B}T>J_{\textrm{H}}$ is applied to mimic this effect. $R_{xx}$
at the four different spin configurations are shown in Fig. \ref{fig:Numerical-simulation-results}c,
where (i), (ii), (iii) and (iv) correspond to the band structures
in Figs. \ref{fig:The-theoretical-figure2}a-d, respectively. We found
that the $R_{xx}$ values of (i) and (iii) are higher than (ii) and
lower than (iv) in this scenario, suggesting the $R_{xx}$ of the
two anti-parallel configurations of $\boldsymbol{S}_{\textrm{T}}$
and $\boldsymbol{S}_{\textrm{B}}$, i.e. ($\mathbf{S}_{\textrm{T}},\mathbf{S}_{\textrm{B}}$)=($\uparrow$,$\downarrow$)
and ($\mathbf{S}_{\textrm{T}},\mathbf{S}_{\textrm{B}}$)=($\downarrow$,$\uparrow$),
are greater and smaller than the $R_{xx}$ in the parallel cases,
respectively. Thus, configuration (ii) produces a dip in $R_{xx}$,
while configuration (iv) produces a peak (Fig. \ref{fig:The-theoretical-figure2}e).
\begin{figure}[h]
\begin{centering}
\includegraphics[width=0.9\columnwidth]{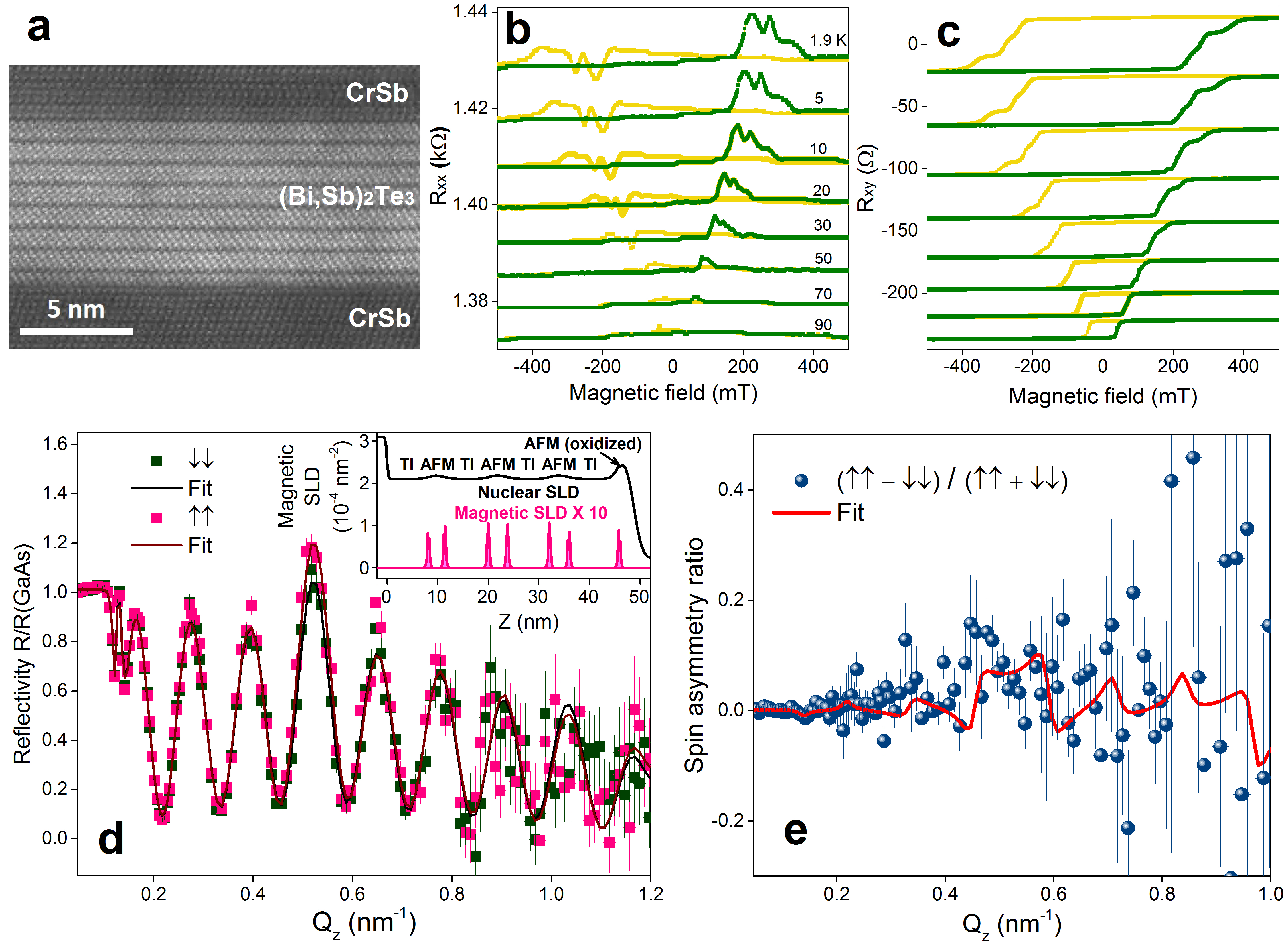} 
\par\end{centering}
\caption{\textbf{Multi-channel magnetic proximity effect in a $\textrm{\textbf{(TI/AFM}}\textrm{\textbf{)}}_{\textbf{n}}$
superlattice (SL).} \textbf{a}, a cross-sectional scanning transmission
electron microscope image obtained from a SL, which captures the sharp
interfaces between the TI layer sandwiched by two AFM layers. There
is a small in-plane rotation between the top and the bottom CrSb layers
when interfacing with TI surfaces, i.e. the top AFM layer is taken
along the ($1\bar{2}10$) zone axis while the bottom one slightly
deviates from it. \textbf{b} and \textbf{c}, $R_{xx}$ and $R_{xy}$
results of the superlattice as functions of external perpendicular
magnetic fields at different temperatures. \textbf{d}, polarized neutron
reflectivity and \textbf{e}, spin asymmetry (at $5\thinspace\textrm{K}$
with a $700\thinspace\textrm{mT}$ in-plane field) normalized to the
GaAs substrate for the spin-polarized $R_{\uparrow\uparrow}$ and
$R_{\downarrow\downarrow}$ channels of the SL. The inset in \textbf{d}
shows the corresponding model with structural and magnetic scattering
length densities (SLDs) used to obtain the best fit. The error bars
are $\pm1$ standard deviation.\label{fig:PNR_results}}
\end{figure}

Following the magnetization switching-related topological transition
model described above, it is expected that the proximity-induced AHE
and the antisymmetric MR features will become more pronounced in the
$\textrm{(AFM/TI)}_{n}$ ($n$ is defined as the repeating number
of the AFM/TI bilayer) superlattice structure where more non-identical
AFM/TI interfaces become involved in the proximity process. To confirm
this scenario, we prepared a high-quality superlattice of $n=4$ (Fig.
\ref{fig:PNR_results}a), and performed the temperature-dependent
magneto-transport measurements. As expected, even stronger antisymmetric
MR and AHE signals are present; furthermore, multiple spikes are observed
in the MR loops during the magnetization reversal, which again coincide
with the multiple Hall-resistance steps near the coercivities. Similar
to the AFM/TI/AFM trilayer case, these features likely indicate the
unsynchronized switchings of different interfacial channels and the
multiple-switching behavior may stem from larger variations in the
switching fields due to the increased number of slightly different
AFM/TI interfaces. Figs. \ref{fig:PNR_results}b and c provide direct
experimental evidences that the magnetization reversal process occurs
in a step-by-step manner in such a superlattice structure. 

To quantitatively investigate the net magnetic polarization with detailed
depth-dependent information, we probed a representative $n=4$ superlattice
using polarized neutron reflectometry (PNR). PNR was performed at
5 K in an applied in-plane magnetic field of $700\thinspace\textrm{mT}$,
and the non spin-flip specular reflectivities, sensitive to the depth
profile of the nuclear scattering length density and net in-plane
magnetization, were measured as a function of the momentum transfer
vector $Q_{Z}$. Fig. \ref{fig:PNR_results}d shows the fitted reflectivities
alongside the most likely nuclear and magnetic depth profile, while
the magnetic features may be more clearly seen in the spin asymmetry
ratio (defined as the difference between the two non spin-flip reflectivities
normalized by their sum) plotted in Fig. \ref{fig:PNR_results}e.
In this figure, with the spin asymmetry, we observe small but nonzero
splitting (a small dip) between the reflectivities near the critical
edge ($Q_{Z}=0.15\thinspace\textrm{nm}^{-1}$) and at the first order
superlattice reflection ($Q_{Z}=0.4-0.6\thinspace\textrm{nm}^{-1}$).
Although the resulting magnetization is extremely weak, the statistical
significance of the two features is three and six standard deviations,
respectively.

In fitting the data, we find several models which may describe the
data. The most plausible of these, shown in the inset of Fig. \ref{fig:PNR_results}d,
exhibits a net magnetization at the interfacial regime. In such a
model, the interfacial TI possesses a net magnetization of approximately
$30\thinspace\textrm{emu}\cdot\textrm{cm}^{-3}$. We note that the
PNR data can also be described quite well by alternative models as
discussed in Supplementary Information. However, the model in which
all of the magnetization is localized within the TI yield splitting
at the superlattice peak which is the opposite sign of the measurement.
Furthermore, confining all of the magnetization uniformly within the
AFM layer cannot simultaneously match the splitting magnitude at the
critical edge and superlattice peak. Finally, although a model assuming
uniform non-zero magnetizations in both the TI and the AFM layers
fits well, there is no clear mechanism by which the bulk TI might
become ferromagnetic through the entire layer. Therefore, all of these
alternative models (Models 1-3 in Supplementary Information) are eliminated
from consideration and we conclude that a modulated magnetic depth
profile must be present, with the magnetization mainly distributed
at the interfaces. This is consistent with the obseved proximity-induced
magnetizations at the AFM/TI interfaces in both the bilayer and trilayer
structures.

In summary, we have experimentally demonstrated transport signatures
which indicate a topological transition induced by the anti-parallel
magnetizations of the top and the bottom surface states in a TI thin
film. The emergent surface magnetization of a TI is induced through
proximity coupling to antiferromagnetic layers. When both surfaces
are magnetized, two intermediate spin configurations, i.e., ($\mathbf{S}_{\textrm{T}},\mathbf{S}_{\textrm{B}}$)=($\uparrow$,$\downarrow$)
and ($\mathbf{S}_{\textrm{T}},\mathbf{S}_{\textrm{B}}$)=($\downarrow$,$\uparrow$),
are introduced during the magnetization reversal process. These intermediate
states restore the overall time-reversal symmetry of the two-surface
system. One of the two intermediate states, i.e. ($\mathbf{S}_{\textrm{T}},\mathbf{S}_{\textrm{B}}$)=($\downarrow$,$\uparrow$),
introduces a set of two Chern numbers of $\pm1$, resulting in two
counter-propagating chiral edge modes inside the exchange gap. This
change in the number of transport channels induces antisymmetric magnetoresistance
spikes during the scan of the external field, which exactly coincide
with the coercive field of the AHE. By growing an $\textrm{(AFM/TI)}_{n=4}$
supperlattice, we introduce more AFM/TI interfaces with different
spin configurations, and can more clearly demonstrate the switching-related
topological transition as well as enhance the AFM-induced proximity
effect. This AFM-TI structure provides a promising new framework for
the manipulation of topological orders in TI thin films.

\section*{Methods}

\subsection{Film growth}

All the films were fabricated using an ultrahigh vacuum molecular
beam epitaxy system. Semi-insulating (resistivity larger than $10^{6}\thinspace\textrm{\textgreek{W}-cm}$\LyXThinSpace )
GaAs (111) substrates were pre-annealed in the growth chamber at 580
\textdegree C for the desorption of the native surface oxide. The
substrate temperature was then reduced and maintained at 200 \textdegree C
for the remained of the fabrication process. High-purity Bi and Cr
were evaporated from standard Knudsen cells while Sb and Te were evaporated
by standard thermal cracker cells. All the epitaxial growth was monitored
by an in-situ reflection high-energy electron diffraction (RHEED).
For the growth of both $\textrm{(Bi,Sb)}_{2}\textrm{Te}_{3}$ and
CrSb layers, the growth conditions were optimized to obtain very sharp,
smooth, streaky RHEED patterns, while the intensity oscillations were
used to calibrate growth rates. These growth rates were determined
to be around 0.08 and 0.06 �/s for $\textrm{(Bi,Sb)}_{2}\textrm{Te}_{3}$
and CrSb, respectively. After growth, 2 nm of Al layers were evaporated
on all the sample surfaces in-situ at room temperature for protection
against contamination and oxidation.

\subsection{Nonequilibrium Green's function calculation}

The NEGF calculation was carried out using the tight-binding model
established from the $k\cdot p$ Hamiltonian (Eq. \ref{eq:K_dot_P_Hamiltonian})
by discretizing the differential operators after applying $k_{\alpha}\rightarrow-i\partial_{\alpha}$.
The terminals were assumed to be semi-infinite leads which were truncated
into self-energy terms: $\boldsymbol{G}^{R}=\left[\epsilon-\boldsymbol{H}_{D}-\sum_{i}\boldsymbol{\Sigma}_{i}\right]^{-1}$,
where $\boldsymbol{G}^{R}$ is the retarded Green's function, $\boldsymbol{H}_{D}$
is the isolated tight-binding Hamiltonian of the device, and $\boldsymbol{\Sigma}_{i}=\boldsymbol{t}_{i}^{\dagger}\boldsymbol{g}_{i}^{R}\boldsymbol{t}_{i}$
is the self energy obtained from the surface Green's function ($\boldsymbol{g}_{i}^{R}$)
and the hopping term between the device and the semi-infinite leads
($\boldsymbol{t}_{i}$). Following the standard Landauer-B�ttiker
linear response picture, the terminal current, $I_{i}$, and the voltage,
$V_{i}=\mu_{i}/e$, can be obtained from $\boldsymbol{I}=\frac{e}{h}\left[\int\left(-\frac{\partial f_{0}}{\partial\epsilon}\right)\boldsymbol{T}\left(\epsilon\right)d\epsilon\right]\delta\boldsymbol{\mu}$,
where $f_{0}\left(\epsilon-\epsilon_{F}\right)$ is the Fermi distribution,
$\boldsymbol{I}=\left(I_{1},I_{2},\cdots,I_{6}\right)^{T}$ denotes
the current at different terminals, and $\delta\boldsymbol{\mu}=\left(\delta\boldsymbol{\mu}_{1},\cdots,\delta\boldsymbol{\mu}_{6}\right)^{T}$
is the chemical potential shift ($\delta\mu_{i}=\mu_{i}-\epsilon_{\textrm{F}}$).
The transmission matrix was obtained by $T_{i,j}=\textrm{Tr}[\boldsymbol{\Gamma}_{i}\mathbf{G}_{i,j}^{R}\boldsymbol{\Gamma}_{j}\mathbf{G}_{i,j}^{A}]$
for $i\neq j$ and $T_{i,i}=\text{\textminus}\sum_{j\neq i}T_{i,j}$,
where $\boldsymbol{\Gamma}_{i}=i\left(\boldsymbol{\Sigma}_{i}-\boldsymbol{\Sigma}_{i}^{\dagger}\right)$
and $\mathbf{G}_{i,j}^{A}=\mathbf{G}_{i,j}^{R\dagger}$. The Hall
resistance can be obtained by fixing the bias along the longitudinal
direction and solving for the bias at different terminals, assuming
zero terminal currents in the transverse direction.

\subsection{Polarized Neutron Reflectometry}

PNR measurements were performed using the polarized beam reflectometer
(PBR) instrument at the NIST Center for Neutron Research. The sample
was cooled to 5 K in an applied in-plane field of 700 mT. Incident
and scattered neutrons were spin polarized either spin-up ($\uparrow$)
or spin-down ($\downarrow$) with respect to the applied magnetic
field. Two spin-flippers and polarizing supermirrors were used to
select the incident and scattered neutron polarization direction.
Thus, we measured the non spin-flip reflectivities as a function of
the momentum transfer along the film normal direction in a range of
$0.1\sim1.0\thinspace\textrm{nm}^{-1}$. The neutron propagation direction
was perpendicular to both the sample surface and the applied field
direction. Based on magnetometry measurements showing in-plane magnetization
saturation below the applied field of 700 mT, spin-flip scattering
is not expected. Therefore, we measured the spin-up and spin-down
reflectivities using full polarization analysis to ensure that the
incident and scattered beams retained identical neutron polarization
directions. We refer, therefore, only to the spin-up and spin-down
non-spin-flip reflectivities, which are functions of the nuclear and
magnetic scattering length density profiles. The magnetic profile
was deduced through modeling of the data with the NIST Refl1D software
package.
\begin{acknowledgments}
The transport measurement and the theoretical modeling in this work
are supported by the Spins and Heat in Nanoscale Electronic Systems
(SHINES), an Energy Frontier Research Center funded by the US Department
of Energy (DOE), Office of Science, Basic Energy Sciences (BES) under
award \#SC0012670. We are also grateful to the support from the National
Science Foundation (DMR-1411085), and the ARO program under contract
W911NF-15-1-10561. We also acknowledge the support from the FAME Center,
one of six centres of STARnet, a Semiconductor Research Corporation
program sponsored by MARCO and DARPA. Certain commercial equipment,
instruments, or materials are identified in this paper to foster understanding.
Such identification does not imply recommendation or endorsement by
the National Institute of Standards and Technology, nor does it imply
that the materials or equipment identified are necessarily the best
available for the purpose.
\end{acknowledgments}

\bibliographystyle{apsrevNoURL}

\end{document}